\documentclass[sigconf]{acmart}
\AtBeginDocument{%
  }

\copyrightyear{2025}
\acmYear{2025}
\setcopyright{rightsretained}
\acmConference[AAR 2025]{The sixth decennial Aarhus conference: Computing X Crisis}{August 18--22, 2025}{Aarhus N, Denmark}
\acmBooktitle{The sixth decennial Aarhus conference: Computing X Crisis (AAR 2025), August 18--22, 2025, Aarhus N, Denmark}
\acmPrice{}
\acmDOI{10.1145/3744169.3744171}
\acmISBN{979-8-4007-2003-1/25/08}

\begin{document}

\title{Towards Creating Infrastructures for Values and Ethics Work in the Production of Software Technologies}

\author{Richmond Y. Wong}
\email{rwong34@gatech.edu}
\orcid{0000-0001-8613-0380}
\affiliation{%
  \institution{Georgia Institute of Technology}
  \city{Atlanta}
  \state{Georgia}
  \country{USA}
}

\renewcommand{\shortauthors}{Wong}

\begin{abstract}
  Recognizing how technical systems can embody social values or cause harms, human-computer interaction (HCI) research often approaches addressing values and ethics in design by creating tools to help tech workers integrate social values into the design of products. While useful, these approaches usually do not consider the politics embedded in the broader processes, organizations, social systems, and governance structures that affect the types of actions that tech workers can take to address values and ethics. This paper argues that creating infrastructures to support values and ethics work, rather than tools, is an approach that takes these broader processes into account and opens them up for (re)design. Drawing on prior research conceptualizing infrastructures from science \& technology studies and media studies, this paper outlines conceptual insights from infrastructures studies that open up new tactics for HCI researchers and designers seeking to support values and ethics in design.  
\end{abstract}




\maketitle

\section{Introduction and a Vignette}
Computing appears to be in a crisis when addressing the ethical and social effects of computing systems. Recognizing how computer systems can embed and promote social values \cite{nissenbaum_how_2001}, human-computer interaction (HCI) and other computing researchers have created a plethora of tools, frameworks, and methods for taking values and ethics into account during computing design, deployment, and education (e.g., \cite{chance_routledge_2025,Chivukula2021,friedman_survey_2017,gray_scaffolding_2023,lee_landscape_2021,wong_seeing_2023}). At the same time, computing development and deployment continues to cause real and potential harms across areas including misinformation \cite{hoover_eating_2023,merken_ai_2025}, privacy and surveillance \cite{Harwell2019a}, workplace equity \cite{Heath2016,Rowland2019}, social biases \cite{pierson_lawsuit_2023,singer_many_2019}, or security \cite{goodin_hacker_2024}. Emerging (generative) AI systems pose a range of new risks to human (and non-human) actors \cite{bender_dangers_2021,shelby_sociotechnical_2023}. 

HCI’s dominant approach to addressing values and ethics is to design tools to help technology developers and designers integrate consideration of social values into systems during the product development and design process. While useful, these approaches tend to assume that individual technology developers and designers have the power to make alternate product design decisions. These usually do not consider the broader processes, organizations, social and economic systems, and governance structures that affect how design is done in the first place and constrain what individual technology workers can do. Growing practice-based research on values and ethics in design has shown the limitations of this design approach, highlighting how barriers to addressing values and ethics in organizational practice stem from: tensions between individual designers’ and organizational practices \cite{ali_walking_2023,gray_ethical_2019,wong_tactics_2021}; how ethics principles are operationalized and align (or do not align) with organizational contexts \cite{madaio_tinker_2024}; a perceived lack of organizational support for responsible design practices \cite{madaio_assessing_2022}; varying amounts of social power \cite{scheuerman_products_2024,widder_its_2023}; how technical discussions are shaped by business priorities \cite{Metcalf2019,ali_walking_2023}; or the invisible labor required to enact value centered and ethical design practices \cite{popova_who_2024,Su2021,rattay_why_2025}. The predominant tool-based approach that HCI has taken does not address these challenges, as exemplified by the following vignette from a prior project conducted by the author investigating how UX professionals attempt to address values and ethics in their work. 

\begin{quote} 
In 2018, I was sitting in a coffee shop in San Francisco. I was meeting Francine, who works on user experience at a large company that creates and sells enterprise software as one of its products. Francine began to discuss an incident where she and other co-workers learned that one of their company’s clients was involved in perpetuating harms against migrant families attempting to cross international borders. That client organization reached out to ask for help to improve their installation of the software made by Francine’s company. Francine and a co-worker had strong feelings against helping this client. They drafted a letter that they planned to share, noting how this violated their personal values. While Francine felt her immediate manager was supportive, the issue got raised to upper management. Francine recounted the response back from a chief officer of the company as basically “do your job,” and that not working for this client would “open a can of worms”---that it might lead to a situation where anyone could stop work based on their personal values. In the end, management at Francine’s company ended up hiring an outside contractor to help this particular client. While Francine was glad she did not personally have to help this client; she was frustrated that this contracting-out solution did not address the underlying harms and concerns that she and her co-workers had voiced. 
\end{quote}

One of the striking things for me from this vignette is that Francine seems to have done the supposed “correct” things that are implicitly expected in HCI’s dominant approaches to addressing values and ethics in design. She identified an ethical issue and voiced her concerns to try to create some type of change. But in the end, noting really changed. While the potential magnitude of harm in this situation may be high compared to other ethical issues, Francine’s account is reflective of research that has since described the social power barriers that tech workers encounter when they try to surface potential ethical issues or harms (e.g., \cite{ali_walking_2023,lev_ari_strategies_2024,madaio_tinker_2024,scheuerman_products_2024,widder_its_2023,wong_tactics_2021}). 

These accounts point to the limits of the dominant HCI approach to values and ethics: helping individual tech workers to identify and address potential ethical issues is not sufficient to actually create and enact ethical change in large corporate contexts. Individual workers often do not have decision making power, and decisionmakers are often motivated by other (often financial) incentives. Thus, helping tech workers like Francine to design ethical systems requires more than creating technical values and ethics tools for individuals. \textbf{This paper seeks to identify approaches to help create the conditions for tech workers to successfully advocate for values and ethics in organizational and corporate contexts during processes of technology production. }

This requires (1) \textit{expanding the landscape of places where interventions by HCI researchers and designers could take place beyond product design practices}, such as in organizational processes, community building among values and ethics advocates, or in law and policy. These processes are intertwined; interventions in one area may open up or foreclose opportunities elsewhere. Learning to operate in this broader landscape may require (2) \textit{new "modes of action"} \cite{mulligan_concept_2020}, referring to how different components of a sociotechnical system act on and interact with each other in different ways. HCI’s predominant mode of action---designing artifacts (such as values and ethics tools) that are evaluated with users---may not always be the most effective mode of action when attempting to create changes across this broader landscape. Working to create changes in organizational practices or law and policy may require different forms of HCI research and design. 

This paper utilizes the lens of \textbf{infrastructures}---a concept from science \& technology studies and media studies that focuses on the social, political, and technical arrangements that allow systems to function---to open up and explore this landscape through multiple modes of action. The paper uses infrastructures in two ways: first, as an analytical lens to help identify new sites and moments of intervention within the sociotechnical infrastructures involved in technology production; secondly, as a mode of action, by suggesting that these sociotechnical infrastructures can be (re)designed in ways that can help improve the conditions for tech workers to advocate for values and ethics in organizational contexts. This paper focuses predominantly on production practices of computing technologies,\footnote{While values and ethics are also shaped and contested by impacted stakeholders during deployment and use (e.g., \cite{harrington_deconstructing_2019,krafft_action-oriented_2021,zong_data_2023}), this paper focuses on how to create conditions that can support values and ethics advocates during technology production processes, following calls to study and intervene in the politics of technology production \cite{posner_agile_2022,gurses_privacy_2017}.}  mostly drawing on examples from a U.S. business context.\footnote{This focus acknowledges the outsized financial impact, power, and forms of control created through the US corporations’ technology development, as part of what Hung describes as a data economy that is best viewed as “planetary assemblages of coloniality” \cite{hung_artificial_2024}.}  

In the following sections, the paper defines key terms and reviews related work on values and ethics. It then presents key insights from infrastructure studies, paired with tactics for research and practice, before outlining new directions for HCI research and design. 

\section{Background and Related Work}

\subsection{Values and Ethics Terminology}
This paper draws on Friedman to define “values” as the guiding principles that people perceive to be broadly desirable, worthwhile, important, or good in life \cite{friedman_value_2008},\footnote{While additional research in HCI and design has further debated the definition of "values" (e.g., \cite{cheng_developing_2010,Shilton2014,LeDantec2009values,jafarinaimi_values_2015}), the Friedman definition suffices for this paper.} and Shilton to define “ethics” as an intertwined but narrower term that refers specifically to moral values \cite{shilton_values_2018}. Given the intertwining of these terms, as well as the common use of both terms in HCI literature in often interchangeable ways, this paper will generally refer to “values and ethics” together to encompass these meanings. The paper then makes use of several related terms:

\begin{itemize}
    \item \textbf{Values and ethics work }draws on my prior definition of values work \cite{wong_tactics_2021} to refer to the everyday practices conducted in the name of attending to social values and ethics, often conducted by tech workers.
    \item \textbf{Values and ethics advocates} draws on Shilton’s concept of a values advocate \cite{shilton_values_2013,shilton_technology_2010}, referring to someone who does values and ethics work by having an interest in and lobbying for social and ethical concerns during a design process or within an organization. In this paper, advocates includes both those with formal roles or responsibilities for addressing values and ethics (e.g., \cite{ali_walking_2023,Metcalf2019}), and those who have other roles but informally advocate or take on extra work to address values and ethics (e.g., \cite{wang_designing_2023,wong_tactics_2021}). 
    \item \textbf{Values and ethics tools} refer to the range of toolkits, tools, design materials, and approaches that attempt to help designers and practitioners consider values and ethics during different phases of the design process, such as those documented in \cite{Chivukula2021,friedman_survey_2017, wong_seeing_2023}.
\end{itemize}

Additionally, when using the term “\textbf{tech workers},” this paper focuses on the workers who are often the imagined users of HCI’s values and ethics tools. These are usually workers who are imagined as taking on the role of values and ethics advocates (formally or informally) during the development of computing systems \cite{abebe_roles_2020,shilton_technology_2010}, which could include, but is not limited to, software engineers, user experience practitioners, user researchers, product managers, responsible AI practitioners, or privacy/security/accessibility specialists. While additional research highlights the broader forms of tech labor required in the developing of computing systems---such as distributed data platform work or janitorial work within tech company offices \cite{gray_ghost_2019,irani_hidden_2016,raval_interrupting_2021,zlolniski_labor_2003}---this paper largely limits its investigation to the narrower group of tech workers who are the imagined users of HCI values and ethics tools. Future work may look to this broader definition of “tech workers” to find and build new solidarities, new sites and moments of intervention, and new modes of action.

\subsection{Addressing Values and Ethics in HCI}
Within computing research, HCI and adjacent fields have designed many values and ethics tools to help tech workers consider and address values and ethical issues during the technology design process, which prior researchers have inventoried and analyzed across multiple dimensions (e.g., \cite{Chivukula2021,friedman_survey_2017,lee_landscape_2021,wong_seeing_2023}). These tools are often evaluated for their usability in laboratory or classroom contexts. 

However, practice-based research on values and ethics work shows that much of the work of doing ethics is social rather than technical. Social and organizational factors affect values and ethics work in practice, including: “levers” that can spur consideration of values and ethics in the design process \cite{shilton_values_2013}, tensions between individual and organizational practices \cite{gray_ethical_2019,wong_tactics_2021}; practitioners’ backgrounds and competencies \cite{chivukula_identity_2021,chivukula_dimensions_2020,scheuerman_products_2024}; practitioners’ information seeking practices \cite{kou_distinctions_2018,kou_what_2018}; rhetorical practices to communicate with organizational decisionmakers \cite{rose_arguing_2016}; interpreting legal requirements \cite{grover_encoding_2024}; the role of community and collective action \cite{pillai_exploring_2022,Spitzberg2020}; and  collaboration across diverse stakeholders and roles such as product managers or cross-functional teams \cite{deng_2023_investigating,lev_ari_strategies_2024}. 

A growing area of research is specific to values and ethics work as it relates to AI systems (often under the banner of “responsible AI”), investigating social and organizational factors such as: tech workers’ views on their ethical responsibility \cite{widder_dislocated_2023}, how AI ethics principles are operationalized and align (or do not) with organizational contexts \cite{rakova_where_2021,schiff_principles_2020}, collaboration among diverse AI ethics teams \cite{deng_2023_investigating}, a perceived lack of organizational support for responsible AI \cite{madaio_assessing_2022,madaio_codesigning_2020,widder_dislocated_2023}, and how technical discussions are shaped by business priorities \cite{passi_making_2020,passi_trust_2018}. 

Thus many challenges in addressing values and ethics are not due to a lack of tools, but rather due to social and organizational factors. While existing values and ethics tools can be useful, these tools usually do not address the social and organizational factors that affect practitioners’ abilities to conduct values and ethics work. Taking these factors into consideration, or potentially changing these factors, may require a different set of lenses and approaches. This paper proposes concepts from infrastructure studies as one such approach. 

\subsection{Infrastructures Studies}
Analytically, this paper draws on the theoretical concept of infrastructures from science \& technology studies and media studies to understand the ongoing sociotechnical assemblages, relationships, and practices needed to maintain systems \cite{parks_signal_2015,star_ethnography_1999,star_steps_1994}. Infrastructures contain specific technologies; are enabled by social institutions through activities such as standards-setting, maintenance, and repair; and support particular human actions while complicating others \cite{bowker_toward_2010,jackson_understanding_2007,star_steps_1994,Wong2020}. 

For example, the internet infrastructure consists of physical cables and data centers, but also consists of global standards and their multi-stakeholder governance processes that ensure interoperability  \cite{doty_internet_2013}, as well as laws and regulations about wired and wireless internet usage \cite{wong_wireless_2015}. At the same time, access to and experiences of internet infrastructure vary widely, due to factors such as economic decisions that limit availability in some regions \cite{Burrell2018}, or specific and diverse cultural histories that shaped the development of networking systems in different ways over time \cite{starosielski_undersea_2015}. Infrastructures are inherently sociotechnical, and help draw attention to the interconnections among the technical, sociocultural, organizational, and political aspects of a system. 

Several qualities of infrastructure as an analytical concept make it useful for considering values and ethics work. First, Bowker and Star advocate that an infrastructural lens asks researchers to conduct an “inversion,” focusing their attention on the relationships, processes, people, and practices that normally exist in the background of a situation or activity \cite{bowker_information_1994,Bowker2000}. With this lens, the social, organizational, political, and economic processes that are often in the background when developing values and ethics  tools are foregrounded as sites of investigation and potential intervention. Second, infrastructures highlight complex assemblages. Bowker et al. write that “Infrastructure is indeed a fundamentally relational concept; it emerges for people in practice, connected to activities and structures” \cite{bowker_toward_2010}. This highlights the interdependencies among the social, organizational, political, and economic processes that affect values and ethics work. Third, infrastructural relationships bridge scale and time: decisions made in the design and creation of infrastructures can have broad impact \cite{irani_critical_2014,spitzberg_creating_2023}. Even though infrastructures change over time and local experiences of the same infrastructure can differ, initial choices in the design of an infrastructure can reverberate long after those decisions were made \cite{jackson_understanding_2007}. Considering values and ethics interventions at an infrastructural level, rather than at the product design level, may enable positive action at a larger scale.  Last, infrastructures can draw attention to systems of power. Star advocates that studies of infrastructures should identify dominant narratives by identifying “that which has been made other, or unnamed,” as well as surfacing work practices that may normally be invisible \cite{star_ethnography_1999}. This can help draw attention to how infrastructural systems affect the social power, agency, and positionalities of values and ethics advocates. 

Computing research has adopted these perspectives on infrastructures in multiple ways. One approach is analytical, using infrastructures as a way to explore and understand sociotechnical systems and the broader contexts in which they operate. This approach highlights maintenance and repair work, standards and policy, and political and economic conditions as objects of study (e.g., \cite{fox_managerial_2019,Houston2016,jackson_rethinking_2014,jackson_breakdown_2014,lindtner_tinkering_2017,steinhardt_breaking_2016,yang_future_2024}). A second approach is to use infrastructures in a more design- or action-oriented way, or as Jackson et al. describe,  as a “sensibility: a way of thinking and acting in the world capable of moving between the separate registers of technical and social action” \cite{jackson_understanding_2007}. Examples  include using concepts from infrastructures studies to help design large information systems \cite{hanseth_theorizing_2008,pipek_infrastructuring_2009}, to consider the long-term maintenance of design projects \cite{irani_critical_2014}, and as lenses to inform speculative design explorations \cite{cheon_amazon_2024,Wong2020}. 

This paper utilizes infrastructures in both an analytical and action-oriented way. The next section highlights specific insights from infrastructure studies that can help researchers analyze the landscape of values and ethics work in new ways, and suggests tactics for how HCI research on values and ethics in design might re-orient or expand its work based on those insights. 

\section{Tactics for Creating Infrastructures for Values and Ethics Work}
This section draws on insights from infrastructures studies to present a set of new tactics for HCI to support values and ethics work.\footnote{The organization of this discussion is similar to Wong et al.’s discussion on infrastructure studies’ potential to influence speculative design practice \cite{Wong2020}.}  Across the tactics, the paper highlights projects in and beyond HCI that are already starting to utilize some of these approaches. By bringing examples these together I hope to suggest a more explicit and concerted effort toward addressing values and ethics utilizing an infrastructural lens. 

Each subsection is grounded in a particular insight from infrastructure studies: (1) infrastructures are dynamic processes; (2) infrastructures are social and experienced from multiple subject positions; (3) infrastructures require ongoing forms of (invisible) work; and (4) infrastructures are embedded in existing systems, calling attention to standards, law, and policy. 

Each subsection describes how that infrastructural insight has affinities with values and ethics work. Following each insight, the paper presents one or more tactics for HCI research and design, illustrated with exemplars and highlighting the modes of action that the tactic helps open up. 

\subsection{Insight: Infrastructures are Dynamic Processes}

Infrastructures are dynamic systems. Bowker et al. consider infrastructures as a set of practices and processes, using the term “infrastucturing” \cite{bowker_toward_2010}. Star and Ruhleder describe that an infrastructural lens askes “when” is an infrastructure, rather than “what” is an infrastructure, focusing on the dynamic relationships and processes that form infrastructures \cite{star_steps_1994}. Exemplifying this, Starosielski investigates undersea internet cable infrastructure in the Pacific Ocean not as a “thing,” but as a set of situated cultural histories, subject to different politics of ownership, regulation, and contestation, through political, environmental, and cultural processes \cite{starosielski_undersea_2015}. 

HCI’s values and ethics tools tend to focus on changing the static outcomes of software design. Once a set of social values or ethical goals are identified, those can be inscribed into the design of the technical system. However, viewing software design in organizational contexts as a set of dynamic processes rather than a set of outcomes opens the door to considering how to design and shape those organizational processes themselves. 

\subsubsection{Tactic: Shifting From Values in Product Design to Values in Process Design}

Like the design of technical artifacts, processes can also embody their own values and politics. Research from administrative law and governance highlights the importance of “procedural values” when designing processes—social values that help stakeholders view processes as legitimate, such as auditability, participation, or transparency of decision-making \cite{doty_internet_2013,freeman_private_2000,koops_criteria_2008}. These are different than “substantive values,” or social values that should be promoted by the outcomes of a process. Prior research on values and ethics work argues that ethical technology design requires organizational processes which support ethical cultures \cite{ali_walking_2023,holstein_improving_2019,madaio_assessing_2022,rakova_where_2021,wong_tactics_2021}. 

Technology companies’ organizational processes embody their own social values, often oriented around market logics \cite{ali_walking_2023,Metcalf2019}. Existing organizational processes include product privacy and security reviews, or user research processes, or various forms of risk and impact assessments. Thus, this tactic asks: what could it look like to re-design these organizational processes? 

One mode of action enact this tactic is to \textbf{influence shared standards and frameworks that articulate what should occur in technology development processes.} For instance, the International Organization for Standardization (ISO) has a standard for “human-centered design for interactive systems,” outlining the process of human centered design \cite{international_organization_for_standardization_iso_iso_2019}. The US National Institute of Standards and Technology’s (NIST) AI Risk Management Framework provides a high level framework of different risk assessment processes that can be used when developing and deploying AI systems, and describes the range of stakeholders that could be included in these processes \cite{national_institute_of_standards_and_technology_nist_artificial_2023}. 

Another mode of action is to \textbf{apply existing value-centered design approaches to designing the values embedded in organizational processes, rather than values embedded in technical artifacts.} HCI researchers and designers can work to change the values embedded in workplace processes, in part by building on a history of (re)designing workplaces, particularly in Scandinavian participatory design \cite{asaro_transforming_2000,gregory_scandinavian_2003}. For instance, Gray et al.’s project to help technology practitioners co-design ethics-focused action plans highlights an approach to how practices can be designed \cite{gray_building_2024}. Expanding this approach to the design of organizational processes could expand researchers' focus from thinking about individual actions to collective and organizational-scale actions. Considerations when working in this mode of action include thinking about how organizational processes embody various procedural values, as well as understanding how different organizational stakeholders may care about or conceptualize procedural values in different ways. For instance, how can decisions be contested? Who gets to participate on what terms? What parts of the process are visible and transparent, and to who? What tensions exist between stakeholders’ perspectives on these processes \cite{gallardo_interdisciplinary_2024}?

A third mode of action can focus on \textbf{re-designing processes that exist beyond the scope of a single organization}. HCI researchers and designers might consider re-designing processes that change the broader contexts that technology production takes place in. For instance, Wong et al. identify financial investment practices in technology companies as a set of processes where values can be surfaced and contested \cite{wong_privacy_2023}. Changing the norms or incentives around technology company financial investment processes might shift how values and ethics get addressed. For instance, this ight include re-defining how environment, social, and governance (ESG) investing accounts for those factors when making investment decisions, or changing the regulations around publicly-traded companies’ disclosure processes and what information they need to publicly disclose about their actions. What if privacy or AI risks of a company's software products were measured and disclosed to investors, or factored into ESG investment practices? Additionally, Widder’s conceptualization of AI production as a “supply chain” highlights additional processes and relationships that occur “upstream” during software production \cite{widder_dislocated_2023}, which could serve as additional points of intervention to re-imagine organizational processes across multiple organizations involved in the supply chain.

HCI researchers and designers, and values and ethics advocates do not have to take the organizational processes and broader economic contexts that shape technology product as given. By conducting an infrastructural inversion and bringing organizational processes to the foreground, these broader processes could be new sites for intervention and (re)design.

\subsection{Insight: Infrastructures are Social and Experienced from Multiple Subject Positions}

While the term “infrastructure” may conjure ideas of large physical systems like pipes and bridges, infrastructure also requires human and social interaction. Lee et al. identify the importance of “human infrastructure,” calling attention to social groups and collaborative practices as a key part of infrastructure \cite{lee_human_2006}. Star and Ruhleder also describe infrastructure as being learned as part of membership in a community, and being linked with the community’s conventions \cite{star_steps_1994}, drawing on Lave and Wenger’s concept of a community of practice in situated learning \cite{lave_situated_1991}. 

Numerous studies of values and ethics work support this perspective by identifying how tech workers: navigate their professional communities and organizational workplace communities \cite{ali_walking_2023,lev_ari_strategies_2024,madaio_assessing_2022,scheuerman_products_2024,wong_tactics_2021}; take on different roles to educate others \cite{chivukula_identity_2021}; and take on or assign different levels of responsibility for conducting values work \cite{popova_who_2024}. While many values and ethics tools focus on helping improve the design of products, fewer focus on these social communities as a point of intervention. 

Because infrastructures are social, they are embedded in systems of power and are experienced in unequal ways by people with different subject positions. For example, Burrell highlights how the visibility of infrastructure is not always in the background based on people’s subject position; for residents in rural areas of the U.S. with scarce or unreliable internet access, the internet infrastructure is quite visible and present \cite{Burrell2018}.

Values and ethics tools often imagine an idealized designer or engineer who has the power to make “better” design decisions after using the tool. However, this does not recognize the subject positions or relations of power that tech workers experience, affecting their ability to make decisions or even use those tools in the first place.

\subsubsection{Tactic: Design to Build Communities of Practice for Values and Ethics Work}

One mode of action to enact this tactic is to \textbf{educate and design in ways that help tech workers navigate the social dynamics of their organizations and build communities of practice that can help support them in doing values and ethics work}. Ethics education modules or design tools might focus on improving people’s human and social skills to help them build social networks and find allies within and beyond their organization. Cha et al.’s “Ethics Pathways” activity helps people reflect on the organizational resources they have to take ethical action, including social and communal resources (as well as technical ones) in an academic research context \cite{cha_ethics_2024}, but could be extended to a corporate tech worker context. However, while that activity helps people identify their resource needs, the tool does not explicitly help build or create those resources. 

Widder et al. recruited teams from companies to play an ethics game that asks players to identify potential harms of AI applications, finding that the use of the game in an industry setting is unlikely to lead to direct changes in products, but may be more helpful at helping players find critically-aligned allies “from which a broader collective to raise critique may be fashioned” \cite{widder_power_2024}. This was a secondary effect of the game, but opens up a new design space: to explicitly design activities or tools that focus on these social tasks like finding allies and building communities of practice. In an example of designing for communities of practice for values advocacy beyond tech workers, Krafft et al.’s “Algorithmic Equity Toolkit,” is aimed at building communities of citizens who can take political action when local governments adopt algorithmic tools \cite{krafft_action-oriented_2021}. Similar tools could be developed for a tech worker context.

In deploying this tactic, lessons may also be drawn from efforts across HCI to design for collective action, including what Le Dantec calls “social design”---designing with and for collective communities in ways that emphasize supporting social action or activism \cite{le_dantec_design_2016}---or Ehn’s view of participatory design as “infrastructuring” community building \cite{ehn_participation_2008}, or design justice’s goals to sustain, heal, and empower communities \cite{costanza-chock_design_2020,design_justice_network_design_2018}. In these, the goal of design is to build community capacities. Lessons may also be learned from efforts within HCI to design for labor organizing efforts \cite{khovanskaya_tools_2019,khovanskaya_bottom-up_2020,wolf_designing_2022}, as well as existing collective action and activist efforts in the tech industry \cite{tan_unlikely_2024,Bhuiyan2019}. Additional modes of action to build communities of practice may take forms beyond design, such as doing community work and outreach, alliance building, or developing and using educational materials. 

\subsubsection{Tactic: Acknowledging Positionalities of Values and Ethics Advocates}

Prior research on values and ethics work highlights the effects of social power and positionality. Scheuerman and Brubaker describe how tech workers’ positionality---including affinities with group identities, job role and function, and positions within teams and within the organization—affect how they recognize and seek to address ethical issues such as computer bias \cite{scheuerman_products_2024}. Hoffmann describes how corporate initiatives to increase diversity and inclusion can end up perpetuating further harms against people from groups that are trying to be included \cite{hoffmann_terms_2020}; when at the same time people from groups traditionally marginalized in the technology industry workforce are often asked to do extra work related to addressing values and ethics. Even though many values and ethics tools contain calls to be “participatory” with diverse stakeholders from inside and outside of the companies developing software technologies, Delgado et al. highlight many ways that “participation” can be enacted, some of which uphold exist power structures and limit the voices of these stakeholders \cite{delgado_participatory_2023}. Widder et al. investigate why engineers may not feel empowered to advocate for values and ethical change—ranging from financial and immigration precarity, to feeling unsafe in a workplace environment, to being sidelined due to organizational incentives \cite{widder_its_2023}. This is not unique to corporate contexts; Horgan and Dourish discuss how data teams in a government context act as “tempered radicals” (drawing on Meyerson and Scully) within their institutional context, upholding certain institutional norms while finding space to challenge others \cite{horgan_ambiguity_2018}.

\textbf{The design and development of values and ethics tools---HCI's main mode of action---might do more to consider the positionalities of the values and ethics advocates using the tools.} Considerations for this mode of action include: Does the intended user have the position and power to advocate for the types of changes that the tool suggests? Is it safe for those advocates to do so? If not, then researchers may determine that other modes of action would better support values and ethics advocates, such as attempting to change organizational processes or standards that might help shift their position and power within the organization. Considering positionalities and the safety of values and ethics advocates suggests that sometimes not increasing the visibility of their work may better help them continue their advocacy. 

\subsection{Insight: Infrastructures Require Ongoing Forms of (Invisible) Work}

Infrastructures do not exist on their own indefinitely; they require ongoing maintenance and repair. Jackson suggests considering the world through the lens of breakdown rather than growth, novelty, or progress, bringing renewed attention to practices of maintenance and repair \cite{jackson_rethinking_2014}. Houston et al. suggest that these moments of repair are also sites and moments where the social values of a system can be surfaced, debated, and reconsidered \cite{Houston2016}. As Star describes, often much of this necessary maintenance and repair work is invisible and operates in the background \cite{star_ethnography_1999}. An infrastructures lens brings these work practices into the foreground for consideration.

Working to address values and ethics in technology production requires forms of work and labor beyond the technical work of designing systems. Wang et al. discuss the additional and less visible forms of work that UX practitioners do to contribute to responsible AI initiatives, such as translating ethical guidelines and frameworks to UX practice, often taking time and energy beyond their formal job requirements \cite{wang_designing_2023}. Madaio et al. describe the articulation work required to take abstracted AI ethics tools (such as an AI fairness checklist) and contextualize them into their own organizational context, such as customizing the tool, integrating it into existing workflows, and navigating tensions in who gets to make these decisions \cite{madaio_codesigning_2020}. Several projects have highlighted the emotional labor conducted by values and ethics advocates, including the effort people undertake to raise concerns while still being viewed as a “good” colleague who does not disrupt too much \cite{wong_using_2021}, feeling vulnerability when taking on responsibility to address ethical issues \cite{popova_who_2024}, or navigating the cognitive and emotional dissonance between tech workers’ experiences inside companies and the external public critiques of those companies \cite{Su2021}. Emotional labor in values and ethics work extends beyond the corporate context, as Rattay et al. find in a European public sector organization, highlighting the affective experiences of failure and moral stress faced by engineers seeking to address values and ethics in their system design \cite{rattay_why_2025}.  Moreover, this labor is not evenly distributed, and often falls on those who have been historically underrepresented in the tech industry \cite{hoffmann_terms_2020}.

\subsubsection{Tactic: Designing for the People Doing Multiple Forms of Values and Ethics Work}

HCI design and research can respond to this in several ways. When using HCI's primary mode of action of designing tools, \textbf{values and ethics tools can more explicitly account for and support these forms of invisible work}. Tools might be designed to gracefully hand off responsibility from an abstracted tool to the human workers doing translation and articulation work to make those tools function in local contexts. For instance, Madaio et al. suggest supporting “positive ambiguity,” that can help tech workers contextualize responsible AI tools, intentionally leaving certain aspects of the tool under-specified (such as example use cases) to allow workers to fill in those details based on their local situated context \cite{madaio_tinker_2024}. Wong et al. suggest that toolkits might explicitly provide guidance on how tech workers can translate and articulate the results of a values and ethics tool into language that would be legible to other organizational stakeholders \cite{wong_seeing_2023}.

Tool design from other subfields may also serve as useful examples. Pierce et al.’s analysis of user-oriented cybersecurity toolkits note several security toolkits that not only provide technical digital security advice, but also describe how to address one’s social and emotional health when protecting their security, acknowledging the emotional labor required to protect security, and how security might be addressed from a social or collective perspective rather than a purely individualistic one \cite{pierce_differential_2018}. Similar forms of guidance could be beneficial in the design of values and ethics tools.

Beyond tool design, other modes of action can support values and ethics work. \textbf{Educating students to build the capacity to do these forms of non-technical work} could be useful in a curriculum context. \textbf{Formal indicators of work, such as organizational performance evaluations, could be changed to make some of this invisible work more visible when that is useful}. Standardized processes (such as a product privacy or security review) might be created that explicitly recognize and support these forms of invisible work. Note that Star and Strauss discuss the complicated ecologies of invisible and visible work; that more visibility can also lead to increased surveillance \cite{star_layers_1999}. Efforts utilizing this tactic should consider the tradeoffs of how and when work is made visible, and consider ways to support the people doing invisible work that may not necessarily make their work more visible. Values and ethics advocates doing work that is in tension with dominant power structures may not want their work to be made visible to decisionmakers, as it may lead to consequences to their reputation or employment (e.g., \cite{metz_second_2021,Metz2020}).\footnote{Particularly in a U.S. context, where few legal labor protections exist.}  For these cases, an alternative tactic or mode of action may be more appropriate.

\subsection{Insight: Infrastructures are Embedded in Existing Systems, Calling Attention to Standards, Law, and Policy}

Infrastructures studies highlights that technologies and artifacts do not exist on their own, but are relationally embedded into broader sociotechnical systems, contexts, and ecologies that affect their creation, adoption, and ongoing use. In contrast, values and ethics tools have been critiqued for being designed in ways that are decontextualized from the complex social and organizational contexts in which those tools might be used and deployed \cite{wong_seeing_2023}, creating an “abstraction trap” \cite{selbst_fairness_2019}. These contextual complexities include tensions between individual and organizational practices at technology companies \cite{gray_ethical_2019}, or the challenges of interdisciplinary collaboration in industry contexts where people across different positions and roles have different views about what constitutes a values or ethical problem, and what the scope of potential solutions might be \cite{deng_2023_investigating,scheuerman_products_2024}. These real-world complexities are often left unmentioned and unaddressed in values and ethics tools. Values and ethics work in industry contexts are also embedded within broader systems that embody technological determinist logics and/or capitalist market logics, which may challenge or even capture attempts to address values and ethics \cite{Metcalf2019,young_confronting_2022,zuboff_age_2019,Phan2021economies,greene_better_2019}.

Infrastructure studies calls particular attention to the role of standards as a way of ensuring that diverse sociotechnical systems can work together \cite{star_steps_1994}. Additional research has expanded this to discuss the role of law and policy in shaping infrastructures \cite{jackson_understanding_2007}, and to discuss law and policy as a type of infrastructure itself \cite{denardis_governance_2016}. Standards and policy can enable changes at scale with far reaching effects, often for relatively long periods of time (though standards and policies can be changed over time as well).

Within HCI, particularly CSCW, policy has been conceptualized as being inherently intertwined with design and practice \cite{jackson_policy_2014}, reflecting the embeddedness of policy as an infrastructure. Policies can both constrain practices and generatively enable new ones, and can occur at multiple levels ranging from national public policy to organizational policy to platform policy \cite{centivany_policy_2016,centivany_popcorn_2016,jackson_why_2013,wong_wireless_2015}. At the same time, standards and policymaking is not a singular force. It has its own values, politics, and histories. Who is able to participate on what terms, how decisions are made and contested, and how policy “on the books” is enacted “on the ground” all vary across contexts \cite{bamberger_privacy_2015,centivany_popcorn_2016,doty_internet_2013,koops_criteria_2008}. Furthermore, while public policymaking may often aspire to democratic ideals, even publicly accountable legal systems can create and reinforce injustices \cite{tran_doing_2024}, particularly when policies are implemented via technical means \cite{eubanks_automating_2015}. 

While a growing number of HCI papers describe “implications for policy” \cite{van_berkel_implications_2023}, few engage in depth with how to specifically implement those policy changes, as that is traditionally viewed as outside the scope of mainstream HCI research \cite{yang_future_2024}.  However, as legal scholars and researchers have turned their attention to the design of technologies as another tool to implement the moral goals promoted by law over the past three decades \cite{cohen_between_2019,cohen_pervasively_2006,hartzog_privacys_2018,hildebrandt_smart_2015,lessig_what_2006,surden_structural_2007}, perhaps it is time for HCI designers and researchers to likewise turn our attention to law, policy, and standards as another tool to promote the values we wish to see in technologies. 

\subsubsection{Tactic: Designing Standards, Policies, and Law to Enable Values and Ethics Work}
\textbf{Changing standards, policies, and law can help shift the dynamics of the organizational and sociotechnical contexts where values and ethics work takes place.} Ali et al. find that directly advocating to change values and ethics during the design process in a corporate setting may not be an effective point of intervention, as individuals may not have the individual social power to enact change—a problem further exacerbated by the frequency of reorganizations that change the structure and makeup of teams \cite{ali_walking_2023}. Spitzberg, highlighting these problems, argues that creating shared standards for addressing values and ethics can shift responsibility away from individuals towards a broader collective  \cite{spitzberg_creating_2023}. While most values and ethics tools are created for use by individual advocates, HCI researchers and designers might consider designing standards, policies, and laws for values and ethics (or designing tools to help those creating standards, policies, and laws). 

Two considerations might shape how this tactic is deployed. First is considering how the design and language of  standards, policies, and law can also codify a particular expertise or approach to addressing an issue—for instance, does a particular social value become viewed as a legal compliance issue, or an engineering issue, or a user centered design issue? Empowering designers and values advocates on the ground requires having advocates during the policy creation process to shape these rules in ways that align with HCI expertise. 

For example, in April 2024, the U.S. National Institute of Standards and Technology (NIST) released a voluntary AI Risk Management Framework, a voluntary standardized framework to help organizations manage potential AI risks \cite{national_institute_of_standards_and_technology_nist_artificial_2023}. Notably, the framework highlights user experience (UX) experts as important organizational stakeholders who should be included during multiple stages of the AI risk management process, including during AI system design and AI system deployment, specifically calling out “domain experts with expertise in human factors, socio-cultural analysis, and governance” \cite[pg35]{national_institute_of_standards_and_technology_nist_artificial_2023}. For organizations following the NIST AI Risk Management Framework, this provides a potential opening for UX values and ethics advocates to assert their expertise and legitimacy for being involved in decision-making about AI systems. Presumably, HCI advocates were involved in the creation of this framework, to make sure that language was included. The international legal landscape can also create opportunities for tech workers. Grover’s study of compliance practices with the European General Data Protection Regulation (GDPR) focuses primarily on North American developers (from companies of different sizes), finding that the GDPR provides an opportunity for developers to help shape their organizations’ interpretation of what “the spirit of the law” is and how their organization’s privacy compliance practices should be conducted \cite{grover_encoding_2024}. 

At other times, standards, policies, and laws may indirectly promote the work of values and ethics advocates by creating new processes or requirements to “piggyback” on, such as Deng et al.’s observation that the creation of privacy impact assessments (often required by data protection laws or in standard privacy management frameworks) created new opportunities for tech workers to advocate for fairness \cite{deng_2023_investigating}. \textit{How might standards, policies, and organizational processes be designed with values and ethics piggybacking explicitly in mind? }Learning from the NIST and Deng et al. examples, standards, policies, and laws might be designed with metaphorical “handholds” to either explicitly or implicitly support the expertise and work of values and ethics advocates.

Second, HCI researchers and designers, as well as values and ethics advocates, can utilize multiple modes of action to affect these policymaking processes. \textbf{They may engage in more traditional forms of policy creation and engagement}, such as directly participating as experts in standards setting processes \cite{doty_internet_2013}, serving on committees such as the ACM US Technology Policy Committee, or producing evidence for policymakers and building social connections with think tanks and policymakers \cite{spaa_understanding_2019}.\footnote{Figuring out how to make these practices legitimate and legible as contributions to the mainstream HCI community rather than being viewed as additional service labor is still debated, as discussed by Yang et al. \cite{yang_future_2024}.}

\textbf{Designerly modes of action can also be used to engage in policymaking}. Margaret Hagan describes opportunities to apply human centered design techniques to the “design” of law and policy \cite{hagan_legal_2020}. Similarly, legal scholar Cristie Ford argues for an alternate approach to regulation rooted in the value of "respect," that "centers people and their lived experiences" in policymaking, utilizing design practices including human centered design, participatory design, and design justice \cite{ford_regulation_2023}. Lindley et al. show how techniques like speculative design can be used in conjunction with policymaking processes to think through the potential implications that policy choices might have on technology design and production \cite{lindley_anticipating_2017}. These modes of action can be used across multiple levels and forms of policymaking, from international standards, to national laws, to industry norms, to organizational and platform policies.

\subsubsection{Tactic: Designing Values and Ethics Tools to Tactically Engage the Lingua Franca of Existing Sociotechnical Systems}

Returning to the design of values and ethics tools as a mode of action, the infrastructural insights presented in the paper suggest new considerations when creating values and ethics tools that are responsive to other modes of action occurring in other parts of the infrastructural landscape. 

\textbf{One design tactic is to create values and ethics  tools that explicitly recognize the social, organizational, and economic contexts where they will be deployed.} In discussing how to make critically oriented HCI more legible to the normative discourse in the field, Khovanskaya et al. describe tactically engaging the lingua franca—the common forms of communication—of the field, sometimes framing critical projects as engaging (and at times upholding) parts of the dominant discourse, while challenging and critiquing others \cite{khovanskaya_double_2015}. 

In conducting values and ethics work, advocates are already making these moves. Deng et al. describe how AI fairness advocates in companies “piggyback” their work on top of established procedures at their companies, such as privacy impact assessments or quantitative metrics to communicate with other teams \cite{deng_2023_investigating}. My prior investigation of “soft resistance” highlights how UX practitioners advocate for values and ethics by critiquing their companies’ products, but largely in ways that are permissible within existing structures \cite{wong_tactics_2021}.

These approaches could be embodied in the design of values and ethics tools themselves. Nathan et al.’s value scenarios provide an early example of this: drawing on the common lingua franca of creating user scenarios in UX, Nathan et al. embed concepts from critical design and value sensitive design into scenarios in order to explicitly consider issues of values and ethics over time, while still being legible to the dominant discourse as a legitimate practice of creating scenarios \cite{nathan_envisioning_2008}. Wong and Khovanskaya similarly investigate the potential for critically oriented speculative design practices which explicitly consider values and ethics to be adopted within mainstream technology companies by framing them as a continuation of the technology industry’s historical practices of corporate futuring, rather than framing them as critically oriented methods \cite{wong_speculative_2018}. 

Moreover, values and ethics tools should consider whose expertise is being legitimized and empowered---whose lingua franca is embodied in the tool? Does the tool suggest an approach aligned with engineering, UX, product management, legal compliance, or another role? This is a design decision that narrows the scope of what gets considered as legitimate ethical action. This is not necessarily negative, as it is a necessary choice when designing tools. However, these decisions should be made intentionally, taking into consideration how this aligns with the types of expertise that are legitimized and empowered through other modes of action, such as if there are certain types of expertise promoted by an adopted standard or a law.

\textbf{The design of values and ethics tools might also consider the cultural contexts where they will be deployed, particularly when making choices in language, rhetoric, and discourse. }Prior research has investigated how UX designers deploy a range of rhetorical strategies to convince other stakeholders about design decisions \cite{rose_arguing_2016}, which may not always be the same rhetorical framings that UX designers use with each other. Building on this and prior research on corporate-level discourses about social values like privacy \cite{mcdonald_powerful_2021}, Wong et al. suggest that values and ethics advocates consider using language that aligns with corporate decisionmakers’ risk assessments when it is tactically useful—for instance, while a UX researcher might view privacy as important because users care about it, describing the need to address privacy in terms of corporate regulatory risks or public relations risks may better align with company decisionmakers’ concerns \cite{wong_privacy_2023}.

Thus, values and ethics tools could be more explicitly designed to “piggyback” on a range of existing discursive practices, processes, and policies that are already seen as legitimate within the organizational contexts where these tools are intended to be deployed. This can make values and ethics legible to other organizational stakeholders as something worth addressing. 

\section{Discussion: Directions for Creating Infrastructures for Values and Ethics Work}

This section reflects on broader directions for research and design in HCI and computing that can support values and ethics work based on the insights and tactics described in the previous section. 

\subsection{Creating Infrastructures for Values and Ethics Work Requires Multiple, Simultaneous Modes of Action}

Taking an infrastructural perspective on values and ethics in design helps HCI researchers and designers, and values and ethics advocates, by (1) analytically opening up new sites of intervention and (2) suggesting new modes of action, as "a way of thinking and acting in the world capable of moving between the separate registers of technical and social action” \cite{jackson_understanding_2007} which could ultimately change these infrastructures. If the broader goal of the values and ethics in design research community is to promote shared social values or morals in technology design, then multiple modes of action must be used at multiple sites of intervention beyond traditional tool design. Drawing on legal research from Lessig, technology design is just one mode of action that can promote moral goals—other modes of actions include law, markets, and social norms \cite{lessig_what_2006}.

An infrastructural perspective highlights an array of modes of action beyond technology design, across a broader landscape beyond the product design process. These modes of actions are interconnected and can better support each other if multiple modes of action are considered simultaneously. Action in one part of the landscape can open up and foreclose opportunities in another part of this landscape. The design of standards, law, and public policy can create changes at broad levels, and codify particular types of expertise that can affect which types of practitioners have agency over these issues within organizations (for instance are values and ethics framed as engineering problems, or user-centered problems, or legal compliance problems?). The design of organizational policy and processes might be informed by requirements in standards, laws, and public policy (such as the requirement to conduct privacy impact assessments), which can create new opportunities for values and ethics work to “piggyback” onto.  Designing to influence financial and economic processes can similarly affect these processes (such as considering what types of business risks companies may have to disclose, affecting the design of risk assessment processes). Building communities of practice to do values and ethics work can utilize knowledge of these processes to find allies or build strategies for how to frame the legitimacy of values and ethics work. Considering changes to processes such as performance reviews can help increase the legitimacy and visibility of values and ethics work when that is beneficial. 

Notably, modes of actions have different politics and operate differently \cite{mulligan_concept_2020}. For instance, who is accountable to who is different when creating public policy, compared to a company’s privacy impact assessment process, compared to a human centered design process. Who gets to participate on what terms, how decisions are made and contested, and what actions are considered legitimate all differ across these modes of action. 

\textbf{HCI and computing can do more than design. }Working towards an infrastructural approach to values and ethics work envisions HCI researchers and values and ethics advocates who can dynamically shift and move across these modes of actions while recognizing their different politics. They can create and shape policies, governance mechanisms, social systems, community actions, and technologies and tools. 

At the same time, this approach also opens up new opportunities for design, such as designing law and policy \cite{hagan_legal_2020,ford_regulation_2023}. It also opens new opportunities to design for indirect change that can still support new communities of practice \cite{ehn_participation_2008}, such as designing systems that can help values and ethics advocates find allies within an organization (who can then work together to enact change) \cite{widder_power_2024}, or designing systems that help ESG financial investors consider a broader range of risks to humans and non-humans when investing in technology companies. 

When working more broadly across these modes of actions, it will be important for HCI researchers and designers to work with and learn from those who have already been working with these modes across research and practice, such as legal researchers, organizational sociologists, labor organizers, or values and ethics advocates. Recognizing and incentivizing these forms of interdisciplinary collaboration can help support this work. 

\subsection{Navigating Hostile Infrastructures}
Existing sociotechnical infrastructures---such as organizational governance processes, market-based accountability mechanisms, public policies, government systems, and technical systems---may not support certain shared human values or even be explicitly hostile to these goals. In addition to seeing infrastructures as a landscape for potential action, HCI researchers and values and ethics advocates should also recognize the potential challenges and threats posted by existing infrastructures and how to navigate them. This includes recognizing differences in positionalities and vulnerabilities among values and ethics advocates, and considering when and to whom to make values and ethics work visible. While visibility can help legitimize values and ethics work, at other times increased visibility can increase surveillance or unwanted attention. 

Doing values and ethics work in this context may also require learning and teaching strategies for navigating and creating change in large organizations (e.g., \cite{lev_ari_strategies_2024,meyerson_tempered_2007}). This is a different design and research problem than designing tools to help tech workers identify values and ethical issues in product design outcomes. HCI designers and researchers should consider designing tools to support the human and social infrastructures of values and ethics work, attuned to consideration of positionality and power. Lessons from historical and contemporary labor activism and organizing in the technology industry and in other industries can inform these design practices (e.g., \cite{khovanskaya_bottom-up_2020,khovanskaya_data_2019,khovanskaya_tools_2019,tan_unlikely_2024}).

Furthermore, conducting and sharing research about values and ethics work done by tech workers in industrial and corporate contexts poses difficulties, particularly in a US context. Corporate infrastructures such as nondisclosure agreements, corporate secrecy practices, and corporate contracting and compliance practices can make it difficult for a researcher to enter a corporate research site, or for a researcher or tech worker to share information outside of the boundaries of a company \cite{scheuerman_walled_2024}. Additional discussion is required in the research community about strategies for conducting research about values and ethics work given these barriers, or what workarounds may be useful when direct access to these research sites is unavailable (such as interviews or document analysis).

\subsection{Re-Imagining and Re-Designing Infrastructures}

In addition to designing for current (potentially hostile) infrastructures, an infrastructural lens also suggests that these infrastructures can be changed, re-imagined, and re-designed. This provides opportunities to embed the infrastructures with different values and politics. Futuring methods such as speculative design can be utilized to envision new configurations of technology production processes that embody different politics than current processes. 

Crafting these alternate imaginaries of technology production processes should be tied to work to shift these infrastructures in the present. This may require new modes of action, as well as new directions for research and design. Researchers might focus more on designing new social and organizational processes to support values and ethics work, rather than tools that focus on technical design choices.  HCI ethics education may highlight social competencies and “soft” skills required to conduct values and ethics work, emphasizing the need to build alliances to create change rather than assuming that a single well-intentioned individual will be capable of enacting product changes on their own. Broader standards, policies, laws, and financial investment practices might be changed to help create environments and conditions that are more conducive to values and ethics work. A plurality of approaches can help work towards changing existing systems that shape the contexts of technology production. 

The paper has focused primarily on technology development in an American corporate business context, in part due to the global (and often colonial) influence that these practices have \cite{hung_artificial_2024}---though international infrastructures such as standards and laws can affect practices in American corporations. Using the lens of infrastructures, this paper identifies sites and modes of action that can help create the conditions for tech workers to advocate for values and ethics within these contexts during processes of technology production. As alluded to in some of the examples presented, these insights may be further applicable to other sectors, countries, or types of organizations, such as academic research environments \cite{shilton_values_2013,cha_ethics_2024} or public sector organizations \cite{horgan_ambiguity_2018,rattay_why_2025}. I also acknowledge that this set of interventions largely focuses on making changes to and within existing institutions of global capital, without directly seeking to change these broader structures. However, I hope that the infrastructural tactics discussed in this paper still hold the possibility to create new sites and moments for critical action. Given the moment of crisis of computing ethics, I see the tactics discussed in this paper as something that can be implemented in the shorter term, complementing a broader range of longer term efforts to enact change or offer alternate infrastructures of technology design (including those that are more directly adversarial to current configurations of global capital). These efforts may take on different forms across geographies, cultures, and communities \cite{escobar_introduction_2018,illich_tools_1973,irani_postcolonial_2010,philip_postcolonial_2012}. 

\section{Conclusion}
Computing is in an ethics crisis. While HCI and adjacent fields have developed a plethora of values and ethics tools, technology mediated harms continue to occur at seemingly increasing frequency and scale. The dominant approach in HCI to address values and ethics in technology design is to create design tools to surface consideration of values and ethics in the product design process. While these tools often imagine an ideal empowered tech worker who can use those tools and then make an alternate design decision, in practice, individual-led change with values and ethics is very difficult---as evidenced by the vignette of Francine in the introduction. While there are individual values and ethics advocates who work at large technology companies, these individuals may not have much social power within an organization, not be decision makers, be concerned about retribution or social consequences, or face competing organizational priorities. Moreover, corporate technology design in North America occurs within the logics of market capitalism, where the concerns voiced by values and ethics advocates are often not a top priority. A continued focus on tool development alone will not address this ethics crisis, which stems from the broader social, economic, and political contexts in which technology production occurs. 

Turning to insights from infrastructure studies suggests several paths and tactics forward beyond these individual-oriented and tool-focused approaches. Insights from infrastructures promotes a more complex ecological understanding of the organizational and political contexts of values and ethics work, suggesting a broader landscape for interventions, as well as new collective, organizational, economic, or political modes of action to address ethics and values in technology design. (Re)designing these infrastructures can help create the conditions for people to do values and ethics work, potentially in ways that are more sustainable and scalable than individual action alone.

\begin{acks}
Thank you to the anonymous reviewers, Catherine Wieczorek, Vera Khovanskaya, Heidi Biggs, Annabel Rothschild, and the Creating Ethics Infrastructure Lab for their helpful and generous feedback on this work. 
Conversations with Deirdre Mulligan, Inha Cha, Sarah Mathew, and Pooja Casula helped inform and inspire this work. 
Conversations with members of the Georgia Tech Critical Computing Working Group were helpful in shaping ideas in this paper, supported by a grant from the Georgia Tech Institute for People and Technology (IPaT) and the Georgia Tech Research Institute (GTRI).
%Ideas in this paper extend ideas from work with prior collaborators including Deirdre Mulligan, Sarah Fox, Nick Merrill, Michael Madaio, and Qian Yang.   
\end{acks}

\bibliographystyle{ACM-Reference-Format}
\bibliography{aarhus,sample-base}


\end{document}